\documentclass{article} 
\pdfoutput=1 
\usepackage{graphicx}
\usepackage{amsmath, amsfonts, amssymb, mathrsfs}
\usepackage{IEEEtrantools}
\usepackage{multirow}
\usepackage{natbib}
\usepackage[colorlinks=true, allcolors=blue]{hyperref}
\newcommand{\R}{\mathbb{R}}

\newcommand{\ud}{\, \mathrm{d}}
\newcommand{\TI}{\Gamma} 
\newcommand{\SI}{Jm$^{-2}$s$^{-0.5}$K$^{-1}$}
\newcommand{\bplane}{$b$-plane}
\newcommand{\PDF}{{\rm pdf}}

\newcommand{\eventA}{\mathcal{A}}
\newcommand{\U}{\mathcal{U}}

\newcommand{\bu}{\mathbf{u}}

\newcommand{\bk}{\mathbf{j}}

\newcommand{\footremember}[2]{%
    \footnote{#2}
    \newcounter{#1}
    \setcounter{#1}{\value{footnote}}%
}
\newcommand{\footrecall}[1]{%
    \footnotemark[\value{#1}]%
} 

\title{Impact Probability Under Aleatory And Epistemic Uncertainties}
\author{Chiara Tardioli\footremember{strathy}{Department of Mechanical \& Aerospace Engineering, University of Strathclyde, Glasgow, G1~1XJ, UK} \and Davide Farnocchia\footremember{JPL}{Jet Propulsion Laboratory, California Institute of Technology, Pasadena, CA~91109, USA} \and Massimiliano Vasile\footrecall{strathy} \and Steve R. Chesley\footrecall{JPL}}

\begin{document}

\maketitle

\begin{abstract}
We present an approach to estimate an upper bound for the impact probability of a potentially hazardous asteroid when part of the force model depends on unknown parameters whose statistical distribution needs to be assumed.
As case study we consider Apophis' risk assessment for the 2036 and 2068 keyholes based on information available as of 2013. 
Within the framework of epistemic uncertainties, under the independence and non-correlation assumption, we assign parametric families of distributions to the physical properties of Apophis that define the Yarkovsky perturbation and in turn the future orbital evolution of the asteroid.
We find ${\rm IP}\leq 5\times 10^{-5}$ for the 2036 keyhole and ${\rm IP}\leq 1.6\times 10^{-5}$ for the 2068 keyhole. These upper bounds are largely conservative choices due to the rather wide range of statistical distributions that we explored.
\end{abstract}

\section{Introduction}
In risk analysis, uncertainties are generally classified into two categories: aleatory and epistemic.
\emph{Aleatory uncertainty} is an inherent variation associated with the physical system or the environment. It may arise from environmental randomness, variation in space, fluctuations in times, or other system variability. 
Given the repetitiveness of the error, it is generally well quantified with a known probability distribution, and, in principle, it cannot be eliminated by further observational data, although it may be better characterized \citep{kolmogorov2018foundations}. 
Conversely, \emph{epistemic uncertainty} is due to a lack of information about the system or phenomenon under investigation. It may come from a lack of experimental data to characterize new materials and processes, poor understanding of coupled physics phenomena, or a lack of knowledge about the model formulation (see, e.g., \citet{helton1994treatment,oberkampf2002error,helton2004exploration,zio2013literature}).

In the last decades there has been an increasing awareness that classical probability theory is inadequate for the treatment of epistemic uncertainty.
Therefore, different non-probabilistic approaches have been developed: imprecise probability, also known as interval analysis, after \citet{walley1991statistical} and \citet{berger1994overview}; probability bound analysis, which combines probability analysis and interval analysis \citep[see, e.g.,][]{ferson1996different}; Dempster-Shafer theory, which uses Belief and Plausibility functions, in the two forms proposed by \citet{dempster1968upper} and \citet{shafer1976mathematical}; possibility theory, which is a special case of interval analysis and Dempster-Shafer theory \cite[see, e.g.,][]{baudrit2006practical}. For a review on methods of representing uncertainty see, e.g., \citet{zio2013literature}.

Recently, \citet{tardioli2016collision} presented an approach to the design of optimal collision avoidance and re-entry maneuvers under uncertainty.
They considered a dynamical model with six aleatory variables (the three components of the position and velocity vectors of the spacecraft) and four epistemic variables (some model parameters). The uncertainty was propagated through the dynamics and the expectation of an optimal deflection maneuver was computed with two different techniques: one using Belief and Plausibility functions and the other using families of parametric distributions.
The Belief and Plausibility functions use no a-priori assumption on the kind of the distribution but any distribution enclosed between an upper and lower distribution is admissible. Whereas, in the parametric distribution approach the assumption is that the probability density function of the uncertain quantity belongs to a family of known distributions parametrized with unknown parameters. 
Following this work, in this paper, we use a parametric distribution approach to include epistemic uncertainties in the estimation of impact probabilities for potentially hazardous asteroids.

Asteroid (99942) Apophis was discovered by R.A. Tucker, D.J. Tholen, and F. Bernardi at Kitt Peak, Arizona on June 2004 (Minor Planet Supplement 109613). With a minimum orbit intersection distance less than 0.05~au and an absolute magnitude less than 22 (i.e., its diameter is larger than 140~m), Apophis is classified as potentially hazardous asteroid. In December 2004, both Sentry\footnote{https://cneos.jpl.nasa.gov/sentry.} and NEODyS\footnote{https://newton.spacedys.com/neodys.} reported a probability of impact with our planet of a few percent in April 2029 \citep{chesley2006potential}.
Although further observations reduced Apophis' orbital uncertainty and ruled out any impact possibility for 2029, the asteroid remains an interesting object worth investigating.
In fact, because of the scattering effect of the 2029 encounter, even small perturbations to the orbit of Apophis affect subsequent impact predictions \citep{giorgini2008predicting}.

\citet{chesley2006potential} shows that the exact encounter circumstances in 2029 depend on the magnitude of the Yarkovsky effect (order of $10^{-15}$~au/day$^2$).
The Yarkovsky effect is a non-gravitational perturbation due to the anisotropic emission of thermal radiation of Solar System objects that causes a secular drift in the semi-major axis \citep{bottke2006yarkovsky,2015aste.book..509V}.
For some near-Earth asteroids the orbital drift associated with the Yarkovsky effect can be measured from the orbital fit to the observations \citep{farnocchia2013near}. That was the case of (101955) Bennu \citep{chesley2014orbit} and (29075) 1950 DA \citep{farnocchia2014assessment}. When the astrometry data does not allow such a detection a Monte Carlo simulation relying on the best physical model currently available is performed.
Nevertheless, the assumed distributions on the asteroid physical parameters are themselves uncertain and may rely on subjective judgement.
That was the case of Apophis as analyzed by \citet{farnocchia2013yarkovsky}, which we revisit within the epistemic uncertainty framework in this paper.

\section{Apophis risk analysis and Yarkovsky effect}

The geometry of close approaches can be described projecting the relative Earth-asteroid distance on an appropriate plane called the \bplane~or target plane, which is defined as the plane including the center of the Earth and normal to the incoming asymptote of the small body trajectory with respect to the planet during the close encounter \cite[see, e.g.,][]{milani2002asteroid,valsecchi2003resonant}. Conventionally, the coordinates on the \bplane~are called $(\xi,\zeta)$ and are defined so that the projection of the Earth's heliocentric velocity onto the \bplane~defines the negative $\zeta$-axis. The $\zeta$ coordinate tells if the asteroid is early or late for the minimum possible distance encounter. The $\xi$ coordinate is the minimum distance that can be obtained by varying the timing of the encounter. The \bplane~coordinates leading to a resonant impact lie on predictable circles \citep{valsecchi2003resonant}. The intersection between the orbital uncertainty region and a Valsecchi circle is called keyhole \citep{chodas1999orbit} and corresponds to a future impact. Assigning a probability to the uncertain variables involved in the dynamical system, a probability of impact can be estimated.
For a comprehensive review on the \bplane~and the corresponding encounter analysis refer to \citet{farnocchia2019planetary}.

\cite{farnocchia2013yarkovsky} presented an impact risk analysis for asteroid (99942) Apophis.
The authors located 20 keyholes on the 2029 \bplane~and found that the probability density function of the $\zeta$-coordinate was completely driven by the dispersion due to the Yarkovsky effect.
The Yarkovsky perturbation was modeled as a transverse acceleration $A_2/r^2$ where $r$ is the heliocentric distance and $A_2$ is a function of certain physical quantities (diameter, Bond albedo, bulk density, thermal inertia, rotational period, and spin orientation).
Thus, different physical characteristics of the asteroid result in a different Yarkovsky perturbation.
\citet{farnocchia2013yarkovsky} assumed a specific distribution for each of the unknown physical parameters that define the Yarkovsky effect. Then, for each keyhole the impact probability was computed as the product of the keyhole width and the value of the $\zeta$ distribution, evaluated in a point of the keyhole. The keyhole related to April 2068, situated in the core of the $\zeta$ probability density distribution, was found to have the highest risk encounter with an impact probability of three in a million. On the other hand, the keyhole associated to April 2036, situated in the tail of the distribution, had the highest width (about $600$~m) with an impact probability of seven in a billion.

These results were superseded by later astrometry and physical characterization \citep{vokrouhlicky2015yarkovsky,brozovic2018goldstone}.
However, in this work we consider Apophis' 2013 scenario and validate our uncertainty quantification method by computing an upper and lower limit for the impact probabilities related to the 2036 and 2068 keyholes on the 2029 \bplane.

\section{Uncertainty Quantification of the Physical Model}
The Yarkovsky perturbation 
can be approximated as \citep{vokrouhlicky2015yarkovsky}
\begin{equation}
    a_t \simeq \dfrac{4(1-A)}{9}\Phi(1~{\rm au})f(\theta)\cos\gamma
\end{equation}
where $A$ is the Bond albedo, $\Phi(1~{\rm au}) = 3 G_S /(2 D\rho c)$ is the standard radiation force factor at 1~au, $G_S = 1361$~W/m$^2$ is the solar constant, $D$ is the asteroid's diameter, $\rho$ is the bulk density, $\gamma$ is the spin axis obliquity, and $\theta_{rot}$ is the thermal parameter, which depends on the rotation period, spin rate, thermal inertia, thermal emissivity, geometric albedo, and radial distance from the Sun. 
These physical quantities are in general poorly characterised, especially shortly after discovery when there is little or no information regarding the key physical characteristics. Therefore, the uncertainties associated to the physical quantities are epistemic rather than aleatory.
Following \citet{tardioli2016collision}, we employ a parametric distribution approach to treat epistemic uncertainties.

\begin{figure*}[h!]
  \centering
  \includegraphics[width=0.9\textwidth]{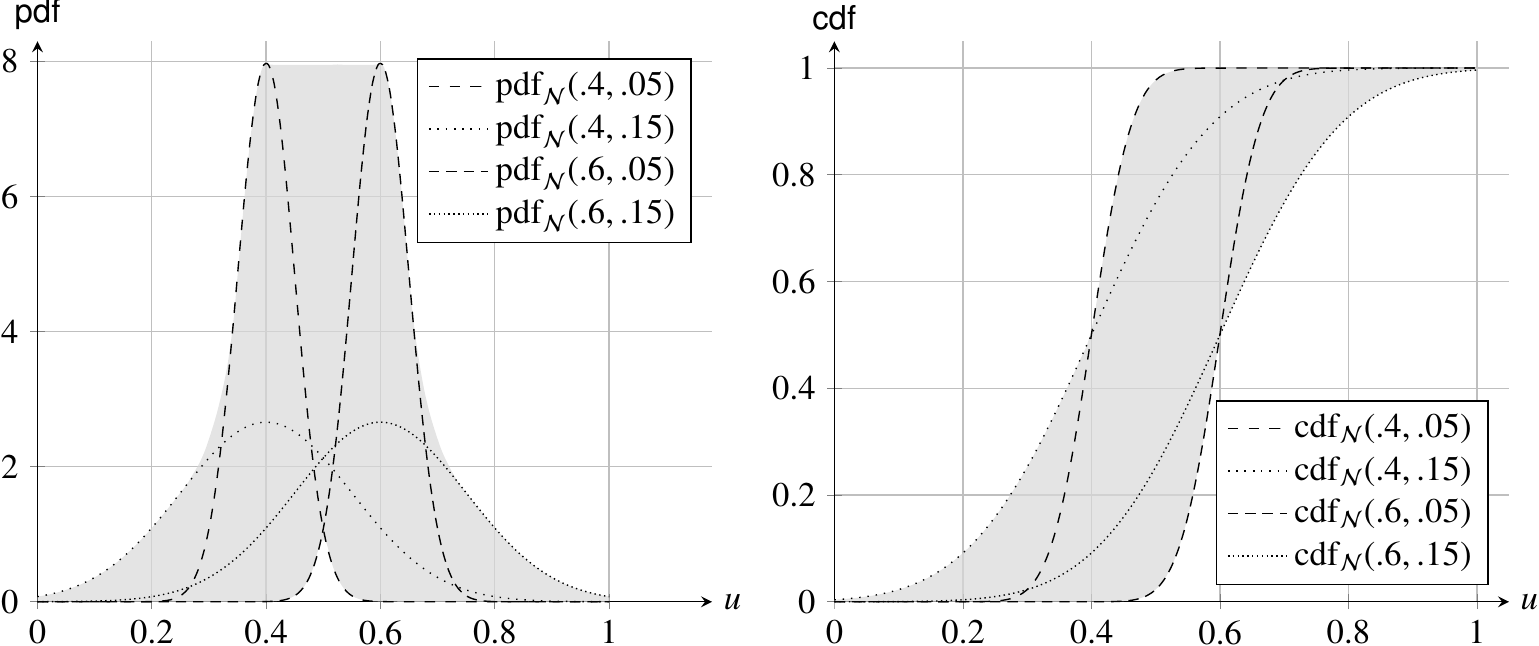}
  \caption{P-box (grey region) defined by a family of Gaussian pdf's (left) and a family of Gaussian cdf's (right).}
  \label{fig:pbox-gauss}
\end{figure*}

In a parametric distribution approach, the actual probability density function is unknown but belongs to a family of distributions parametrized with unknown parameters. 
For example, a family of Gaussian probability density function (${\rm pdf}_\mathcal{N}$) with mean $\mu\in[0.4,0.5]$ and standard deviation $\sigma\in[0.5,1.5]$ is
\begin{equation}\label{eq:gauss-family-pdf}
    [{\rm pdf}_\mathcal{N}] = \{ {\rm pdf}_\mathcal{N}(u;\mu,\sigma):
    0.4\leq \mu\leq 0.6\,,\  0.05\leq \sigma\leq 0.15\,,\ \forall u\in\R\}\,,
\end{equation}
or, using Gaussian cumulative distribution functions (${\rm cdf}_\mathcal{N}$),
\begin{equation}\label{eq:gauss-family-cdf}
    [{\rm cdf}_\mathcal{N}] = \{ {\rm cdf}_\mathcal{N}(u;\mu,\sigma) :
    0.4\leq \mu\leq 0.6\,,\ 0.05\leq \sigma\leq 0.15\,,\ \forall u\in\R\}\ .
\end{equation}
The families of distributions in Eq.~\eqref{eq:gauss-family-pdf} and Eq.\eqref{eq:gauss-family-cdf}, often called probability box, or, shortly, p-box, are illustrated in Fig.~\ref{fig:pbox-gauss}.
A p-box of pdf's (resp. cdf's) is a set enclosing the minimum and maximum values of the pdf's (resp. cdf's). Particularly, the p-box of cdf's can always be enclosed from above and from below by piece-wise curves, called, respectively, upper expectation $E_u$ and lower expectation $E_l$ \citep{ferson2015constructing}:
\begin{equation}\label{eq:pbox-F}
[{\rm cdf}] = \{{\rm cdf}\::\: E_l \leq {\rm cdf} \leq E_u\}\ .
\end{equation}

In the multivariate case, under the independence and non-correlation assumption, the joint probability distribution belongs to a p-box whose elements are the product of univariate pdf's. For example, the p-box of two families of Gaussian pdf's as in Eq.~\eqref{eq:gauss-family-pdf} can be written as
\begin{equation}
    \big[\PDF_{\mathcal{N}_1\mathcal{N}_2}\big] = \big\{\PDF_{\mathcal{N}_1} \times \PDF_{\mathcal{N}_2}\: : \: \PDF_{\mathcal{N}_1}\in\big[\PDF_{\mathcal{N}_1}\big]\ \wedge\ 
    \PDF_{\mathcal{N}_2}\in\big[\PDF_{\mathcal{N}_2}\big]
    \big\}\ .
\end{equation}
A similar definition holds for the product of two families of Gaussian cdf's.

For the \citet{farnocchia2013yarkovsky} scenario, the families of distributions associated to Apophis' physical quantities describing the Yarkovsky effect are reported in Table~\ref{tab:parameter-space}. To enhance the variability in the impact probability, we purposely considered a wide range of parameters, thus not necessarily representing the most realistic scenario.

\begin{itemize}
\item {\em Diameter.} 
\citet{delbo2007albedo} estimate a diameter $D=(270\pm 60)$~m, while \citet{muller2014thermal} derive greater values $D=375^{+14}_{-10}$~m. \citet{licandro2016gtc} confirm a 
diameter in range 380--393~m. We model the diameter uncertainty with normal distributions with mean between $270$ and $385$~m, and standard deviation between $5$ and $60$~m.

\item {\em Slope parameter.} 
\citet{pravec2014tumbling} assume a slope parameter $G=0.24\pm 0.11$ for S and Q types asteroids. 
\citet{mommert2014physical} use $G=0.18\pm 0.13$, obtained averaging all $G$ measurements of asteroids from JPL Small-Body Database Search Engine\footnote{JPL Small-Body Database Search Engine: \url{http://ssd.jpl.nasa.gov/sbdb_query.cgi}.}. \citet{farnocchia2013yarkovsky} assumed $G = 0.15\pm 0.05$. Therefore, we choose normal distributions with mean 
between $0.15$ and $0.24$, and standard deviation between $0.05$ and $0.15$.

\item {\em Geometric albedo.} From polarimetric observations \citet{delbo2007albedo} find a geometric albedo $p_V=0.33\pm 0.08$.
\citet{muller2014thermal} obtain $p_V=0.30^{+0.05}_{-0.06}$ from far-infrared observations with ESA's Herschel space observatory. Recent thermal infrared observations put $p_V$ in the range 0.24--0.33
\citep{licandro2016gtc}, in accordance with radar observations that gives $p_V=0.35\pm 0.10$ \citep{brozovic2018goldstone}. \citet{farnocchia2013yarkovsky} used $p_V = 0.23\pm 0.02$. Therefore, we use lognormal distributions whose corresponding Gaussian distributions have mean between 0.23 and 0.35, and standard deviation between 0.02 and 0.10.
Then, the Bond albedo can be derived from the slope parameter and the geometric albedo through the formula $A=(0.29 + 0.684 G)\,p_V$ \citep{bowell1989application}.

\item {\em Bulk density.} From the grain density and total porosity reported in \citet{binzel2009spectral}, \citet{farnocchia2013yarkovsky} obtained a lognormal distribution with mean $2.2$~g/cm$^3$ and variance $0.3$~g$^2$/cm$^6$. However, Apophis is an Sq-type asteroid and the typical density of this taxonomic class is $2.78 \pm 0.85$~g/cm$^3$ \citep{carry2012density}. Thus, we use lognormal distributions with mean between 2 and 3~g/cm$^3$, and standard deviation between 0.5 and 0.85~g/cm$^3$.

\item {\em Thermal Inertia.} According to
\citep[Fig. 6]{delbo2007thermal} the thermal inertia $\TI$ of near-Earth objects ranges between 100 and 1\,000~\SI, thus the uncertainty of $\TI$ can be represented by $\log_{10}(\TI)=2.5\pm 0.5$. \citet{muller2014thermal} use full range of $\TI$ of 250--800~\SI, with best solution at $\TI= 600$~\SI, giving $\log_{10}(\TI)=2.7\pm0.25$. \citet{licandro2016gtc} constrain the thermal inertia of Apophis to lie in the range 50--500~\SI, giving $\log_{10}(\TI) = 2.2\pm 0.5$.
\citet{farnocchia2013yarkovsky} used a generic relationship between the thermal inertia and the diameter which led to $\log_{10}(\TI)=2.65\pm 0.08$.
Hence, we consider a normal distribution for $\log_{10}(\TI)$ with mean between 2.2 and 2.7, and standard deviation between 0.08 and 0.5.

\item {\em Rotational period.} The uncertainty on the rotational period is very small and it does not affect the impact probability \citep{farnocchia2013yarkovsky}. \citet{pravec2014tumbling} report a rotational period $P_{\rm rot}=(30.56\pm 0.01)$~h, therefore we used a normal distribution with this parameter.

\item {\em Obliquity.}
In our scenario we neglect Apophis' complex rotation and retrograde motion \citep{pravec2014tumbling}.  We consider a simple discrete model with 4 bins in $-1\leq \cos\gamma\leq 1$, and positive frequencies $f_1,f_2,f_3,f_4$ such that $f_1+\ldots +f_4=1$.
\end{itemize}

\begin{table}[t]
\centering
\caption{Distribution families associated to the uncertain variables defining the Yarkovsky effect for Apophis' 2013 scenario. With $\mu$ and $\sigma$ we denote the mean and the standard deviation of the corresponding distribution. Note that the first six variables are epistemic, the last one is aleatory. For units refer to Table~\ref{tab:full-range}.}
\label{tab:parameter-space}
\begin{tabular}{lcccc}
\hline\noalign{\smallskip}
Variable & Distribution Type & Parameter Ranges\\
\noalign{\smallskip}\hline\noalign{\smallskip}
\multirow{2}{*}{Diameter} & \multirow{ 2}{*}{Normal} & $270\leq\mu\leq 385$\\ & & $5\leq \sigma\leq 60$\\ 
& & \\ 
\multirow{2}{*}{Slope parameter} & \multirow{2}{*}{Normal} & $0.15\leq\mu\leq 0.24$\\ & & $0.05\leq \sigma\leq 0.15$\\ & & \\ 
\multirow{2}{*}{Geometric albedo} & \multirow{2}{*}{Lognormal} & $0.23\leq\mu\leq 0.35$\\ & & $0.02\leq \sigma\leq 0.10$\\ 
& & \\ 
\multirow{2}{*}{Bulk density} & \multirow{2}{*}{Lognormal} & $2\leq\mu\leq 3$\\ & & $0.50\leq \sigma\leq 0.85$\\ 
& & \\ 
\multirow{2}{*}{$\log_{10}({\rm Thermal\ Inertia})$} & \multirow{2}{*}{Normal} & $2.2\leq\mu\leq 2.7$\\ & & $0.08\leq \sigma\leq 0.50$\\
& & \\ 
\multirow{2}{*}{Rotational period} & \multirow{ 2}{*}{Normal} & $\mu=30.56$\\ & & $\sigma=0.01$\\ 
& & \\ 
\multirow{4}{*}{Obliquity cosine} & \multirow{4}{*}{4-bin distr.} & $0\leq f_1\leq 1$\\ & & $0\leq f_2\leq 1-f_1$\\ & & $0\leq f_3\leq 1-f_1-f_2$\\ & & $0\leq f_4\leq 1-f_1-f_2-f_3$\\
\noalign{\smallskip}\hline
\end{tabular}
\end{table}

\begin{table}[t]
\centering
\caption{Uncertainty space defined by the physical quantities involved in the Yarkovsky effect for Apophis' 2013 scenario.}
\label{tab:full-range}
\begin{tabular}{lcccc}
\hline\noalign{\smallskip}
Variable & Symbol & Lower Bound & Upper Bound & Unit\\ \noalign{\smallskip}\hline\noalign{\smallskip}
Diameter & D & 50 & 600 & m\\
Slope parameter & $G$ & $-0.424$ & 0.8 & - \\
Geometric albedo & $p_V$ & 0 & 1 & - \\
Bulk density & $\rho$ & 0.5 & 5 & g/cm$^3$ \\
$\log_{10}({\rm Thermal\ Inertia})$ & $\log_{10}(\TI)$ & 1 & 10000 & \SI \\
Rotational period & $P_{\rm rot}$ & 30.5 & 30.6 & h\\
Obliquity cosine & $\cos\gamma$ & $-1$ & 1 & - \\
\noalign{\smallskip}\hline
\end{tabular}
\end{table}

In total, we have six epistemic variables and one aleatory variable. The uncertainty analysis under aleatory and epistemic uncertainty starts from the definition of the uncertainty space $\U$, which is a 7-dimensional hyper-rectangle defined by the full ranges of Apophis' physical parameters involved in the Yarkovsky effect (see Table~\ref{tab:full-range}). The underlying assumption is that of independence and non-correlation among the uncertain variables.

\citet{farnocchia2013yarkovsky} map the Yarkovsky-related semi-major axis drift onto the 2029 $b$-plane and its $\zeta_{\rm 2029}$ coordinate. Denoting this map as $f:\U\to\R$, we want to assess the probability of the event $\eventA=\{u\in\U:|f(u)-\zeta_o|\leq w/2\}$, where $\zeta_o$ is the center of a generic keyhole on the 2029 \bplane~and $w$ its width. If all uncertain variables are aleatory, then the probability of $\eventA$ coincides with the impact probability associated with the keyhole $\zeta_o$: ${\rm IP}={\rm P}(\eventA)$. Instead, if the distribution parameters associated to the uncertain physical quantities vary into intervals, then the probability of $\eventA$ is bound by a lower and upper value given by
\begin{equation}\label{eq:distribution_full}
{\rm P}_*(\eventA) = \min_{\bk\in J}\int_{\eventA} \PDF_{\bk}(\bu)\, \ud \bu\,,\qquad
{\rm P}^*(\eventA) = \max_{\bk\in J}\int_{\eventA} \PDF_{\bk}(\bu)\, \ud \bu\,,
\end{equation}
where $\bu=(D,G,p_V,\rho,\TI,\cos\gamma)$ is the epistemic variable vector,
\begin{equation}
\bk=(\mu_1,\sigma_1,\ldots,\mu_5,\sigma_5,f_1,f_2,f_3)
\end{equation}
is the unknown distribution parameter vector (the parameters defining the first six distribution families in Table~\ref{tab:parameter-space}), $J=J_1\times\ldots J_{13}$ is the distribution parameter space with $J_i$ the range defined in Table~\ref{tab:parameter-space} for each $i=1,\ldots 13$, and $\PDF_\bk$ is the joint probability distribution product varying in the p-box $[\PDF_\bk:\bk\in J]$ with
\begin{IEEEeqnarray}{rCl}
\PDF_\bk &\propto & \PDF_{\mathcal{N}(\mu_1,\sigma_1)}\times \PDF_{\mathcal{N}(\mu_2,\sigma_2)}\times \PDF_{\mathcal{LN}(\mu_3,\sigma_3)}\times \PDF_{\mathcal{LN}(\mu_4,\sigma_4)} \nonumber\\
&&\times\, \PDF_{\mathcal{N}(\mu_5,\sigma_5)}\times
\PDF_{\mathcal{N}(\mu_{rot},\sigma_{rot})}\times
\PDF_{\mathcal{D}_4(f_1,f_2,f_3,1-f_1-f_2-f_3)}\,,
\end{IEEEeqnarray}
where $\mathcal{N}$ is the Gaussian, $\mathcal{LN}$ is the Lognormal, and $\mathcal{D}_4$ the discrete 4-bin distribution. 
As a consequence, the impact probability associated to the keyhole $\zeta_o$ is itself uncertain and the following inequalities hold
\begin{equation}
{\rm IP}_* = {\rm P}_*(\eventA)\leq {\rm IP}\leq{\rm P}^*(\eventA) = {\rm IP}^*\ .
\end{equation}

The two optimisation problems in Eq.~\eqref{eq:distribution_full} can be solved numerically by replacing the calculation of the exact integrals with approximated forms using a sampling scheme.
In fact, given $M$ sample points $\bu_1,\ldots,\bu_M$,
each integral in Eq.~\eqref{eq:distribution_full} can be approximated as
\begin{equation}\label{eq:integral_approx}
  \sum_{k=1}^{M} I_\mathcal{A}(\bu_k)\, p_k\,,\quad p_k\propto\PDF_\bk(\bu_k)\,,\quad \sum_{k=1}^M p_k = 1\,,
\end{equation}
where $I_\mathcal{A}$ is the indicator function of the set $\eventA$, that is 1 if $\bu_k$ belongs to $\eventA$, 0 otherwise. In this work we will use a scheme with $10^{12}$ uniformly distributed Monte Carlo points.

\section{Results}

We consider Apophis' 2013 scenario and the keyholes found by \citet{farnocchia2013yarkovsky} on the 2029 \bplane. The uncertainties for the physical parameters are reported in Tables~\ref{tab:parameter-space} and \ref{tab:full-range}. For two keyholes, we compute the maximum and minimum of the impact probability. One keyhole corresponds to an impact in 2036. This is the widest keyhole and is located in the tail of the probability distribution. The other keyhole corresponds to an impact in 2068. It is much smaller but within the core of the distribution and yielded the highest impact probability in \citet{farnocchia2013yarkovsky}.
These two keyholes require negative values of $A_2$ in order to be possible. In this paper, we are considering all possible distributions for the obliquity (which determines the sign of $A_2$). Thus, the minimum value of IP is always zero and our uncertainty analysis gives an upper bound for IP rather than a proper interval.

\begin{figure*}[p]
\includegraphics[width=\textwidth]{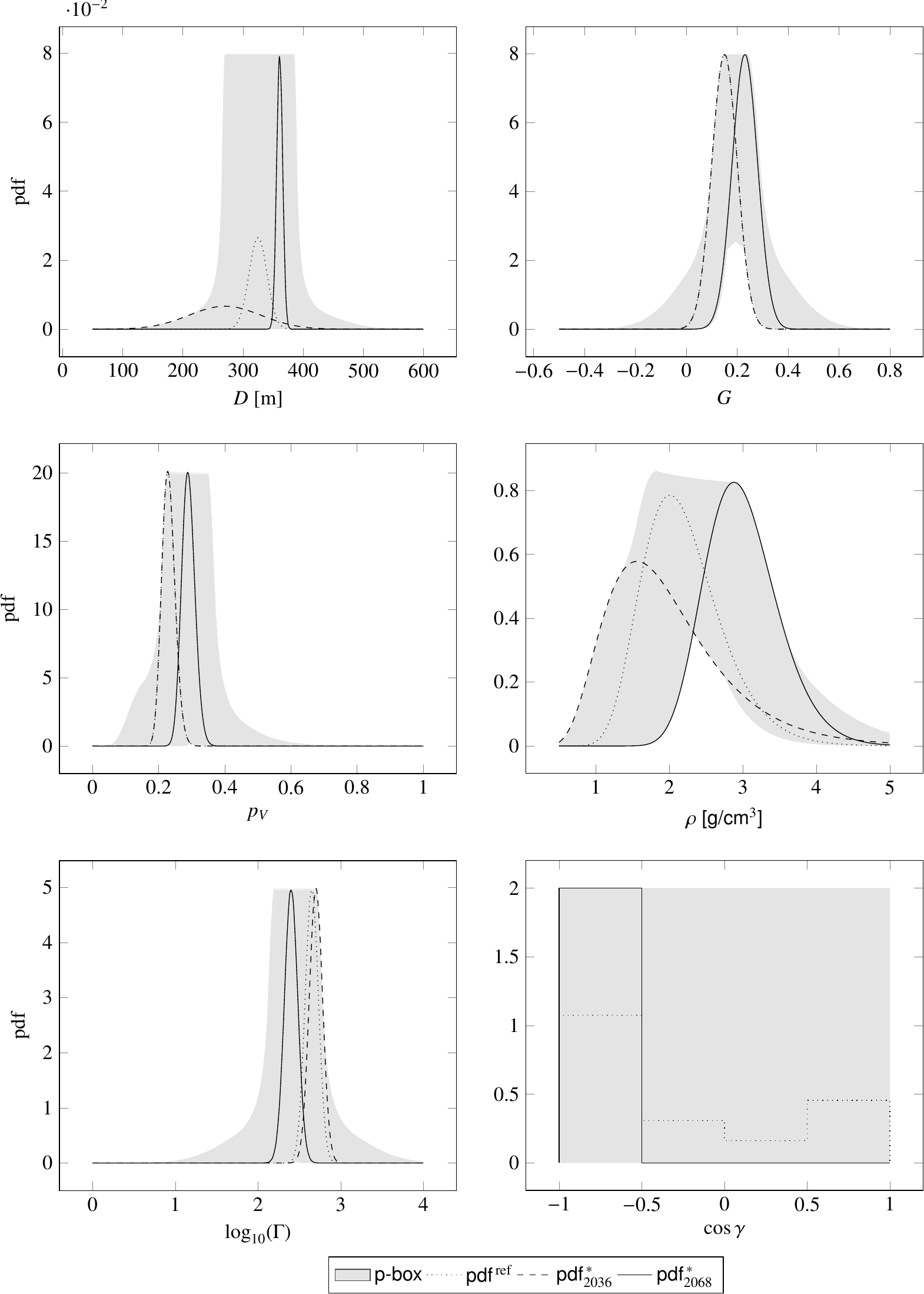}
\caption{Probability boxes and probability density functions of the asteroid physical quantities for the 2036 and 2068 keyholes corresponding to IP$^*$ and IP$^{\rm ref}$.}
\label{fig:pbox-pdf-2068}
\end{figure*}

\begin{figure*}[h!]
  \centering
  \includegraphics[width=0.49\textwidth]{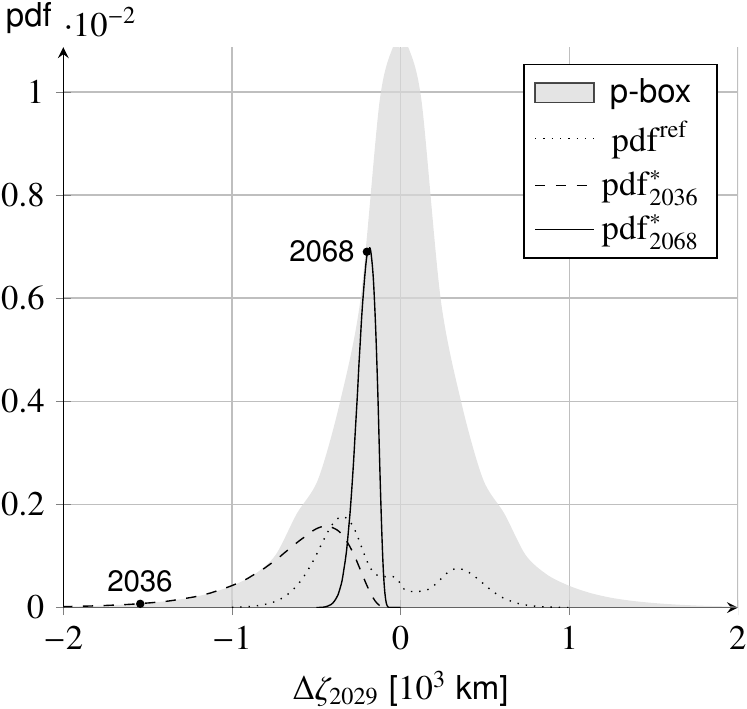}
  \includegraphics[width=0.49\textwidth]{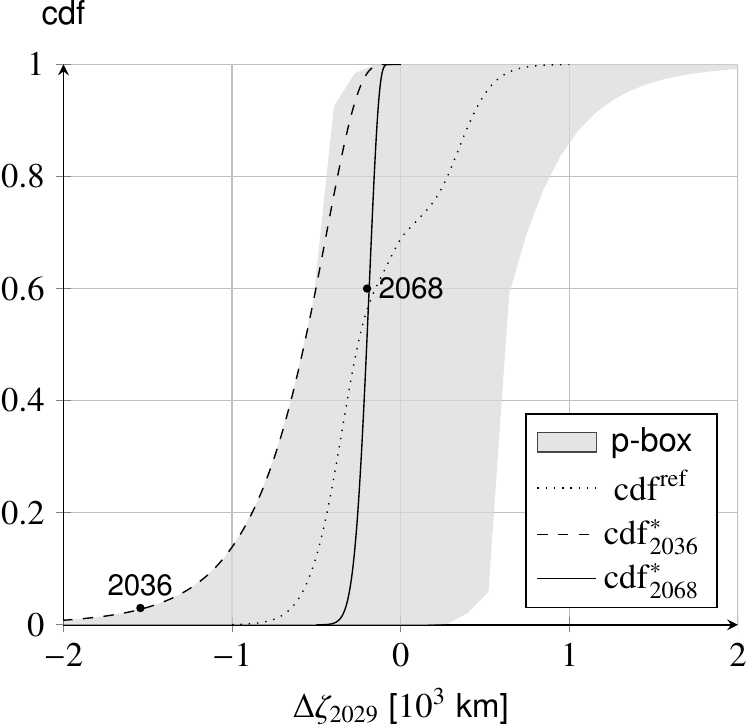}
  \caption{Probability box on the 2029 $b$-plane associated to the distribution families in Table~\ref{tab:parameter-space} in the pdf-space (left) and the cdf-space (right). The origin of the $\zeta_{2029}$-coordinate is set at 47659.24~km. The curves represent the distribution functions corresponding to ${\rm IP}^{\rm ref}$ and IP$^*$ for the 2036 and 2068 keyholes.}
  \label{fig:pbox-zeta}
\end{figure*}

Results are displayed in Table~\ref{tab:IIP}.
To validate our method, for each keyhole, we compute a reference impact probability 
corresponding to fixed distributions derived by the ones used in \citet{farnocchia2013yarkovsky} (see Table~\ref{tab:ref-param}). 
The distribution parameters corresponding to ${\rm IP}^*$ are found using multiple runs of a global optimization method proposed by \citet{Vasile2011Inflationary} and called IDEA (Inflationary Differential Evolution Algorithm). The values, out of all the runs of IDEA, that give the highest impact probability are reported in Table~\ref{tab:optimal-values} and illustrated in Figure~\ref{fig:pbox-pdf-2068}. Figure~\ref{fig:pbox-zeta} shows the uncertainty region of $\zeta_{2029}$ due to the uncertainties on the physical parameters associated to the Yarkovsky effect. The distributions associated to the reference and to the maximum impact probability for the 2036 and 2068 keyholes are also displayed. 

\begin{table}[h!]
    \centering
    \caption{Upper bound of the impact probability (IP$^*$) for the 2036 and 2068 keyholes relative to the distribution families in Table~\ref{tab:parameter-space}. Other columns are: width and position of the center of the keyhole on the 2029~\bplane, impact probability (IP$^{\rm ref}$) corresponding to the reference distribution parameters.}
    \label{tab:IIP}
    \begin{tabular}{cccccc}
        \hline\noalign{\smallskip}
        Year & Width [m] & Position [km] & IP$^{\rm ref}$ & IP$^*$ \\
        \noalign{\smallskip}\hline\noalign{\smallskip}
        2036 & 616.16 & 46115.5 & $5.0\times 10^{-9}$ & $4.6\times 10^{-5}$\\
        2068 & 2.25 & 47447.1 & $2.2\times 10^{-6}$ & $1.6\times 10^{-5}$\\ 
        \noalign{\smallskip}\hline
    \end{tabular}
\end{table}

\begin{table}[h!]
    \centering
    \caption{Reference distribution parameters.}
    \label{tab:ref-param}
    \begin{tabular}{ccc}
        \hline\noalign{\smallskip}
        Variable & Distribution Type & Reference Parameters\\
        \noalign{\smallskip}\hline\noalign{\smallskip}
        $D$ & Normal &  $(325 \pm 15)$~m \\
        $G$ & Normal & $(0.15 \pm 0.05)$\\
        $p_V$ &  Lognormal & $(0.23 \pm 0.02)$\\
        $\rho$ & Lognormal & $(2.2 \pm \sqrt{0.3})$~g/cm$^3$\\
        $\log_{10}(\TI)$ & Normal & $(2.65 \pm 0.08)$\\
        $P_{rot}$ & Normal & $(30.56 \pm 0.01)$~h\\
        $\cos\gamma$ & 4-bin distr. & 53.68\%, 15.44\%, 8.09\%, 22.79\% \\
        \noalign{\smallskip}\hline
    \end{tabular}
\end{table}

\begin{table}[!h]
    \centering
    \caption{Distribution parameters corresponding to the maximum impact probability (IP$^*$) for the 2036 and 2068 keyholes, respectively.}
    \label{tab:optimal-values}
       \begin{tabular}{ccc}
        \hline\noalign{\smallskip}
        Variable & 2036 Max & 2068 Max\\
        \noalign{\smallskip}\hline\noalign{\smallskip}
        $D$ & $(270 \pm 60)$~m &  $(361 \pm 5)$~m \\
        $G$ & $(0.15 \pm 0.05)$ & $(0.23 \pm 0.05)$\\
        $p_V$ & $(0.23 \pm 0.02)$  & $(0.29 \pm 0.02)$\\
        $\rho$ & $(2.0 \pm 0.85)$~g/cm$^3$ & $(3 \pm 0.5)$~g/cm$^3$\\
        $\log_{10}(\TI)$ & $(2.70 \pm 0.08 )$ & $(2.41 \pm 0.08)$\\
        $P_{rot}$ & $(30.56 \pm 0.01)$~h & $(30.56 \pm 0.01)$~h\\
        $\cos\gamma$ & 100\%, 0\%, 0\%, 0\% & 100\%, 0\%, 0\%, 0\% \\
        \noalign{\smallskip}\hline
    \end{tabular}
\end{table}

\section{Discussion and Conclusions}

We presented an approach to include epistemic uncertainty in the physical model of the Yarkovsky effect and to estimate an upper bound for the impact probability of a potentially hazardous asteroid. To illustrate the method we considered asteroid Apophis and the 2036 and 2068 keyholes as discussed by \citet{farnocchia2013yarkovsky}. 
The maximum impact probability required the solution of an optimization problem with a multidimensional integral as objective function. 

Our analysis was not meant to provide the most current and realistic hazard assessment for Apophis. Rather, for demonstration purpose, we chose a wide range of parameters that result in very conservative upper bounds for the impact probability.
For example, if we had fixed the 4-bin distribution of the obliquity and set the frequencies to the reference values, we would have found a maximum impact probability of $2.69\times 10^{-6}$ for the 2036 keyhole and of $7.13\times 10^{-6}$ for the 2068 keyhole.

As more asteroids are discovered, the number of cases where the impact hazard assessment is affected by the Yarkovsky effect and its modeling is set to increase. The technique presented in this paper provides a rigorous framework to characterize the sensitivity of the impact probability to the physical model assumptions and in turn characterize its degree of variability.

The spin state of Apophis may change as a result of the 2029 encounter \citep{scheeres2005abrupt}. As already discussed by \citet{farnocchia2013yarkovsky}, such a change would slightly change the location of the keyholes but have a small effect on the impact probability. Therefore, our analysis remains valid even in that case.
Eventually, we remark that that the parametric distribution approach can be applied to more general families of distributions, for example, defined by a weighted combination of kernels.

\section{Acknowledgements}
We would like to thank the editor AC and the reviewer GV for their recommendations.
This is a post-peer-review, pre-copyedit version of an article published in Celestial Mechanics and Dynamical Astronomy. The final authenticated version is available online at: https://doi.org/10.1007/s10569-020-09991-3

\bibliographystyle{apalike}
\bibliography{paper-IPQ}

\begin{thebibliography}{}

\bibitem[Baudrit and Dubois, 2006]{baudrit2006practical}
Baudrit, C. and Dubois, D. (2006).
\newblock Practical representations of incomplete probabilistic knowledge.
\newblock {\em Computational statistics \& data analysis}, 51(1):86--108.

\bibitem[Berger et~al., 1994]{berger1994overview}
Berger, J.~O., Moreno, E., Pericchi, L.~R., Bayarri, M.~J., Bernardo, J.~M.,
  Cano, J.~A., De~la Horra, J., Mart{\'\i}n, J., R{\'\i}os-Ins{\'u}a, D.,
  Betr{\`o}, B., et~al. (1994).
\newblock An overview of robust bayesian analysis.
\newblock {\em Test}, 3(1):5--124.

\bibitem[{Binzel} et~al., 2009]{binzel2009spectral}
{Binzel}, R.~P., {Rivkin}, A.~S., {Thomas}, C.~A., {Vernazza}, P., {Burbine},
  T.~H., {DeMeo}, F.~E., {Bus}, S.~J., {Tokunaga}, A.~T., and {Birlan}, M.
  (2009).
\newblock {Spectral properties and composition of potentially hazardous
  Asteroid (99942) Apophis}.
\newblock {\em Icarus}, 200:480--485.

\bibitem[Bottke~Jr et~al., 2006]{bottke2006yarkovsky}
Bottke~Jr, W.~F., Vokrouhlick{\'y}, D., Rubincam, D.~P., and Nesvorn{\'y}, D.
  (2006).
\newblock The yarkovsky and yorp effects: Implications for asteroid dynamics.
\newblock {\em Annu. Rev. Earth Planet. Sci.}, 34:157--191.

\bibitem[Bowell et~al., 1989]{bowell1989application}
Bowell, E., Hapke, B., Domingue, D., Lumme, K., Peltoniemi, J., and Harris,
  A.~W. (1989).
\newblock Application of photometric models to asteroids.
\newblock In {\em Asteroids II}, pages 524--556.

\bibitem[{Brozovi{\'c}} et~al., 2018]{brozovic2018goldstone}
{Brozovi{\'c}}, M., {Benner}, L.~A.~M., {McMichael}, J.~G., {Giorgini}, J.~D.,
  {Pravec}, P., {Scheirich}, P., {Magri}, C., {Busch}, M.~W., {Jao}, J.~S.,
  {Lee}, C.~G., {Snedeker}, L.~G., {Silva}, M.~A., {Slade}, M.~A., {Semenov},
  B., {Nolan}, M.~C., {Taylor}, P.~A., {Howell}, E.~S., and {Lawrence}, K.~J.
  (2018).
\newblock {Goldstone and Arecibo radar observations of (99942) Apophis in
  2012-2013}.
\newblock {\em Icarus}, 300:115--128.

\bibitem[Carry, 2012]{carry2012density}
Carry, B. (2012).
\newblock Density of asteroids.
\newblock {\em Planetary and Space Science}, 73(1):98--118.

\bibitem[{Chesley}, 2006]{chesley2006potential}
{Chesley}, S.~R. (2006).
\newblock {Potential impact detection for Near-Earth asteroids: the case of
  99942 Apophis (2004 MN 4 )}.
\newblock In {Daniela}, L., {Sylvio Ferraz}, M., and {Angel}, F.~J., editors,
  {\em Asteroids, Comets, Meteors}, volume 229 of {\em IAU Symposium}, pages
  215--228.

\bibitem[Chesley et~al., 2014]{chesley2014orbit}
Chesley, S.~R., Farnocchia, D., Nolan, M.~C., Vokrouhlick{\'y}, D., Chodas,
  P.~W., Milani, A., Spoto, F., Rozitis, B., Benner, L.~A., Bottke, W.~F.,
  et~al. (2014).
\newblock Orbit and bulk density of the osiris-rex target asteroid (101955)
  bennu.
\newblock {\em Icarus}, 235:5--22.

\bibitem[Chodas, 1999]{chodas1999orbit}
Chodas, P. (1999).
\newblock Orbit uncertainties, keyholes, and collision probabilities.
\newblock In {\em Bulletin of the American Astronomical Society}, volume~31,
  page 1117.

\bibitem[{Delb{\`o}} et~al., 2007a]{delbo2007albedo}
{Delb{\`o}}, M., {Cellino}, A., and {Tedesco}, E.~F. (2007a).
\newblock {Albedo and size determination of potentially hazardous asteroids:
  (99942) Apophis}.
\newblock {\em Icarus}, 188:266--269.

\bibitem[{Delb{\`o}} et~al., 2007b]{delbo2007thermal}
{Delb{\`o}}, M., Dell'Oro, A., Harris, A.~W., Mottola, S., Mueller, M., et~al.
  (2007b).
\newblock Thermal inertia of near-earth asteroids and implications for the
  magnitude of the yarkovsky effect.
\newblock {\em Icarus}, 190(1):236--249.

\bibitem[Dempster et~al., 1968]{dempster1968upper}
Dempster, A.~P. et~al. (1968).
\newblock Upper and lower probabilities generated by a random closed interval.
\newblock {\em The Annals of Mathematical Statistics}, 39(3):957--966.

\bibitem[Farnocchia et~al., 2013a]{farnocchia2013yarkovsky}
Farnocchia, D., Chesley, S., Chodas, P., Micheli, M., Tholen, D., Milani, A.,
  Elliott, G., and Bernardi, F. (2013a).
\newblock Yarkovsky-driven impact risk analysis for asteroid (99942) apophis.
\newblock {\em Icarus}, 224(1):192--200.

\bibitem[Farnocchia et~al., 2013b]{farnocchia2013near}
Farnocchia, D., Chesley, S., Vokrouhlick{\'y}, D., Milani, A., Spoto, F., and
  Bottke, W. (2013b).
\newblock Near earth asteroids with measurable yarkovsky effect.
\newblock {\em Icarus}, 224(1):1--13.

\bibitem[Farnocchia and Chesley, 2014]{farnocchia2014assessment}
Farnocchia, D. and Chesley, S.~R. (2014).
\newblock Assessment of the 2880 impact threat from asteroid (29075) 1950 da.
\newblock {\em Icarus}, 229:321--327.

\bibitem[Farnocchia et~al., 2019]{farnocchia2019planetary}
Farnocchia, D., Eggl, S., Chodas, P.~W., Giorgini, J.~D., and Chesley, S.~R.
  (2019).
\newblock Planetary encounter analysis on the b-plane: a comprehensive
  formulation.
\newblock {\em Celestial Mechanics and Dynamical Astronomy}, 131(8):36.

\bibitem[Ferson and Ginzburg, 1996]{ferson1996different}
Ferson, S. and Ginzburg, L.~R. (1996).
\newblock Different methods are needed to propagate ignorance and variability.
\newblock {\em Reliability Engineering \& System Safety}, 54(2-3):133--144.

\bibitem[Ferson et~al., 2015]{ferson2015constructing}
Ferson, S., Kreinovich, V., Grinzburg, L., Myers, D., and Sentz, K. (2015).
\newblock Constructing probability boxes and dempster-shafer structures.
\newblock Technical report, Sandia National Lab.(SNL-NM), Albuquerque, NM
  (United States).

\bibitem[Giorgini et~al., 2008]{giorgini2008predicting}
Giorgini, J.~D., Benner, L.~A., Ostro, S.~J., Nolan, M.~C., and Busch, M.~W.
  (2008).
\newblock Predicting the earth encounters of (99942) apophis.
\newblock {\em Icarus}, 193(1):1--19.

\bibitem[Helton, 1994]{helton1994treatment}
Helton, J.~C. (1994).
\newblock Treatment of uncertainty in performance assessments for complex
  systems.
\newblock {\em Risk analysis}, 14(4):483--511.

\bibitem[Helton et~al., 2004]{helton2004exploration}
Helton, J.~C., Johnson, J.~D., and Oberkampf, W.~L. (2004).
\newblock An exploration of alternative approaches to the representation of
  uncertainty in model predictions.
\newblock {\em Reliability Engineering \& System Safety}, 85(1-3):39--71.

\bibitem[Kolmogorov and Bharucha-Reid, 2018]{kolmogorov2018foundations}
Kolmogorov, A.~N. and Bharucha-Reid, A.~T. (2018).
\newblock {\em Foundations of the theory of probability: Second English
  Edition}.
\newblock Courier Dover Publications.

\bibitem[{Licandro} et~al., 2016]{licandro2016gtc}
{Licandro}, J., {M{\"u}ller}, T., {Alvarez}, C., {Al{\'{\i}}-Lagoa}, V., and
  {Delbo}, M. (2016).
\newblock Gtc/canaricam observations of (99942) apophis.
\newblock {\em Astronomy \& Astrophysics}, 585:A10.

\bibitem[{Milani} et~al., 2002]{milani2002asteroid}
{Milani}, A., {Chesley}, S.~R., {Chodas}, P.~W., and {Valsecchi}, G.~B. (2002).
\newblock {Asteroid Close Approaches: Analysis and Potential Impact Detection}.
\newblock In {\em Asteroids III}, pages 55--69.

\bibitem[{Mommert} et~al., 2014]{mommert2014physical}
{Mommert}, M., {Farnocchia}, D., {Hora}, J.~L., {Chesley}, S.~R., {Trilling},
  D.~E., {Chodas}, P.~W., {Mueller}, M., {Harris}, A.~W., {Smith}, H.~A., and
  {Fazio}, G.~G. (2014).
\newblock {Physical Properties of Near-Earth Asteroid 2011 MD}.
\newblock {\em The Astrophysical Journal Letters}, 789:L22.

\bibitem[M{\"u}ller et~al., 2014]{muller2014thermal}
M{\"u}ller, T., Kiss, C., Scheirich, P., Pravec, P., O’Rourke, L., Vilenius,
  E., and Altieri, B. (2014).
\newblock Thermal infrared observations of asteroid (99942) apophis with
  herschel.
\newblock {\em Astronomy \& Astrophysics}, 566:A22.

\bibitem[Oberkampf et~al., 2002]{oberkampf2002error}
Oberkampf, W.~L., DeLand, S.~M., Rutherford, B.~M., Diegert, K.~V., and Alvin,
  K.~F. (2002).
\newblock Error and uncertainty in modeling and simulation.
\newblock {\em Reliability Engineering \& System Safety}, 75(3):333--357.

\bibitem[Pravec et~al., 2014]{pravec2014tumbling}
Pravec, P., Scheirich, P., {\v{D}}urech, J., Pollock, J., Ku{\v{s}}nir{\'a}k,
  P., Hornoch, K., Gal{\'a}d, A., Vokrouhlick{\'y}, D., Harris, A., Jehin, E.,
  et~al. (2014).
\newblock The tumbling spin state of (99942) apophis.
\newblock {\em Icarus}, 233:48--60.

\bibitem[Scheeres et~al., 2005]{scheeres2005abrupt}
Scheeres, D., Benner, L., Ostro, S., Rossi, A., Marzari, F., and Washabaugh, P.
  (2005).
\newblock Abrupt alteration of asteroid 2004 mn4's spin state during its 2029
  earth flyby.
\newblock {\em Icarus}, 178(1):281--283.

\bibitem[Shafer, 1976]{shafer1976mathematical}
Shafer, G. (1976).
\newblock {\em A mathematical theory of evidence}, volume~42.
\newblock Princeton university press.

\bibitem[Tardioli and Vasile, 2016]{tardioli2016collision}
Tardioli, C. and Vasile, M. (2016).
\newblock Collision and re-entry analysis under aleatory and epistemic
  uncertainty.
\newblock In Majji, M., { Turner}, J., Wawrzyniak, G., and Cerven, W., editors,
  {\em Astrodynamics 2015}, volume 156 of {\em Advances in Astronautical
  Sciences}, pages 4205--4220. Univelt Inc.

\bibitem[Valsecchi et~al., 2003]{valsecchi2003resonant}
Valsecchi, G.~B., Milani, A., Gronchi, G.~F., and Chesley, S.~R. (2003).
\newblock Resonant returns to close approaches: Analytical theory.
\newblock {\em Astronomy \& Astrophysics}, 408(3):1179--1196.

\bibitem[Vasile et~al., 2011]{Vasile2011Inflationary}
Vasile, M., Minisci, E., and Locatelli, M. (2011).
\newblock An inflationary differential evolution algorithm for space trajectory
  optimization.
\newblock {\em IEEE Transactions on Evolutionary Computation}, 15(2):267--281.

\bibitem[{Vokrouhlick{\'y}} et~al., 2015a]{2015aste.book..509V}
{Vokrouhlick{\'y}}, D., {Bottke}, W.~F., {Chesley}, S.~R., {Scheeres}, D.~J.,
  and {Statler}, T.~S. (2015a).
\newblock {The Yarkovsky and YORP Effects}.
\newblock In {\em Asteroids IV}, pages 509--531.

\bibitem[{Vokrouhlick{\'y}} et~al., 2015b]{vokrouhlicky2015yarkovsky}
{Vokrouhlick{\'y}}, D., Farnocchia, D., {\v{C}}apek, D., Chesley, S.~R.,
  Pravec, P., Scheirich, P., and M{\"u}ller, T.~G. (2015b).
\newblock The yarkovsky effect for 99942 apophis.
\newblock {\em Icarus}, 252:277--283.

\bibitem[Walley, 1991]{walley1991statistical}
Walley, P. (1991).
\newblock Statistical reasoning with imprecise probabilities.

\bibitem[Zio and Pedroni, 2013]{zio2013literature}
Zio, E. and Pedroni, N. (2013).
\newblock {\em Literature review of methods for representing uncertainty}.
\newblock FonCSI.

\end{thebibliography}

\end{document}